\documentclass[sigconf]{aamas}

\usepackage{balance} 
\makeatletter
\gdef\@copyrightpermission{
  \begin{minipage}{0.2\columnwidth}
   \href{https://creativecommons.org/licenses/by/4.0/}{\includegraphics[width=0.90\textwidth]{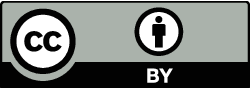}}
  \end{minipage}\hfill
  \begin{minipage}{0.8\columnwidth}
   \href{https://creativecommons.org/licenses/by/4.0/}{This work is licensed under a Creative Commons Attribution International 4.0 License.}
  \end{minipage}
  \vspace{5pt}
}
\makeatother

\setcopyright{ifaamas}
\acmConference[AAMAS '25]{Proc.\@ of the 24th International Conference
on Autonomous Agents and Multiagent Systems (AAMAS 2025)}{May 19 -- 23, 2025}
{Detroit, Michigan, USA}{Y.~Vorobeychik, S.~Das, A.~Nowé  (eds.)}
\copyrightyear{2025}
\acmYear{2025}
\acmDOI{}
\acmPrice{}
\acmISBN{}

\acmSubmissionID{699}

\title{Explaining Facial Expression Recognition}

\author{Sanjeev Nahulanthran}
\affiliation{
  \institution{Monash University}
  \city{Melbourne}
  \country{Australia}}
\email{sanjeev.nahulanthran@monash.edu}

\author{Leimin Tian}
\affiliation{
  \institution{CSIRO Robotics}
  \city{Melbourne}
  \country{Australia}}
\email{leimin.tian@data61.csiro.au}

\author{Dana Kuli\'c}
\affiliation{
  \institution{Monash University}
  \city{Melbourne}
  \country{Australia}}
\email{dana.kulic@monash.edu}

\author{Mor Vered}
\affiliation{
  \institution{Monash University}
  \city{Melbourne}
  \country{Australia}}
\email{mor.vered@monash.edu}

\begin{abstract}
    Facial expression recognition (FER) has emerged as a promising approach to the development of emotion-aware intelligent agents and systems. 
    However, key challenges remain in utilizing FER in real-world contexts, including ensuring user understanding and establishing a suitable level of user trust. We developed a novel explanation method utilizing Facial Action Units (FAUs) to explain the output of a FER model through both textual and visual modalities. We conducted an empirical user study  evaluating user understanding and trust, comparing our approach to state-of-the-art eXplainable AI (XAI) methods.
    Our results indicate that visual AND textual as well as textual-only FAU-based explanations resulted in better user understanding of the FER model.
    We also show that all modalities of FAU-based methods improved appropriate trust of the users towards the FER model.
\end{abstract}


\newcommand{\BibTeX}{\rm B\kern-.05em{\sc i\kern-.025em b}\kern-.08em\TeX}

\begin{document}


\pagestyle{fancy}
\fancyhead{}


\maketitle 


\section{Introduction}

Automatic facial expression recognition (FER) systems, which recognize human emotions from visual information using machine learning approaches, have been proposed to support emotion-aware intelligent systems. FER systems have been adopted in applications in multiple domains such as e-learning~\cite{imani2019survey}, healthcare~\cite{hasnul2021electrocardiogram}, automated driver-assistance~\cite{kolli2011non} and security~\cite{sajjad2019raspberry}. The performance of FER systems has been continuously improving, especially with the advancements in data-driven deep-learning approaches. However, FER systems are far from perfect, due to technical limits and human emotional expressions being personal and context-dependant~\cite{tian2022aac, barrett2019emotional}. This has led to urgent challenges in utilizing FER systems to achieve real-world benefits, as highlighted by the recent discussion of potentially banning the use of AI technologies including FER systems in the EU's AI Act~\cite{edwards2021eu}. One major concern is the lack of transparency and explainability in deep-learning based FER~\cite{linardatos2020explainable}. This hinders an end-user's understanding of these systems as well as their ability to calibrate trust towards the system's outputs and use this technology to their benefit~\cite{abdul2018trends}. 

The growing literature in eXplainable AI (XAI) suggests numerous benefits in providing explanations to end-users, namely in enabling better understanding of AI model predictions~\cite{vilone2020explainable} while mitigating bias~\cite{vered2023effects} and helping users gain greater confidence in model predictions~\cite{van2020interpretable}. However, in FER, the majority of existing research has focused on providing explanations to model developers~\cite{del2022understanding}, instead of end-users. A recent work~\cite{nahulanthran2024ithinkneedhelp} demonstrates the benefits of FER explanations on improving user understanding as well as perceived and demonstrated trust of a FER system. However, as only one XAI method was evaluated in this study, it is unclear which explanation types and presentation modalities in current XAI literature are the most effective for explaining FER to end-users. Moreover, this work did not investigate trust calibration, i.e., whether or not the user's trust towards the FER model aligns with its performance. 
As explanations can influence a user's trust in a system's capabilities, care must be taken to reduce over or under reliance~\cite{miller2022we} which may impair a user's decision making in mission critical settings. 

We therefore investigate the effect of explanations on user trust and understanding of FER systems. We first propose a novel XAI method, named \textit{DEFAULTS}   (\textbf{D}eterministic \textbf{E}xplanations through a \textbf{F}acial \textbf{A}ction \textbf{U}nit visua\textbf{L}-\textbf{T}extual \textbf{S}ystem) that explains FER with visual and textual information grounded in facial expression and emotion theories. We then  measure \textit{appropriate trust}, i.e., when the user's belief in the system aligns with the system's output accuracy, as well as \textit{system understanding}, measured as the alignment between a user's prediction of the FER model's output and the model's actual output.  We empirically show, through a user study that this novel method yields higher appropriate trust and understanding in users compared to state-of-the-art XAI methods.

Using the contribution taxonomy proposed in~\cite{wobbrock2012seven}, our work has the following contributions:
 \textbf{Artifact}: We developed a novel explanation method (\textit{DEFAULTS}) using predicted Facial Action Units (FAUs) to generate a combined textual and visual explanation for FER models. Visual explanations are shown through highlighted facial landmarks associated with activated FAUs, while textual explanations describe the FAU activations; \textbf{Empirical}: We demonstrate that users who are provided with visual and textual FAU-based explanations on FER have a better understanding of the system's predictions. We also demonstrate that this type of explanation method engenders higher appropriate trust in the system as compared to not having explanations; and \textbf{Dataset}: We contribute a dataset\footnote{https://bridges.monash.edu/articles/dataset/DEFAULTS\_Dataset/28443197} containing images, explanations, participant survey and free-text responses for all conditions, collected from a total of 280 participants.

\section{Related Work}

\subsection{Facial Expression Recognition (FER)}

FER is a widely adopted, non-contact method for recognising human  expressions and emotions conveyed by images and facial movements which has garnered growing research  in the past few years~\cite{zeng2007survey, castellano2023automatic}. FER has been shown to be naturalistic,  unobtrusive, economical and easy to deploy and maintain using commercially available sensing systems~\cite{kolli2011non, imani2019survey, kulke2020comparison, mone2015sensing}. As the field continues to advance rapidly, it is crucial to provide a deeper understanding of the underlying machine learning models so that end-users can have a greater sense of autonomy when using these systems~\cite{abdul2018trends}.

Despite recent advancements, several challenges remain in developing accurate, reliable, and robust FER systems that can operate in real-world contexts.  FER literature has focused on improving classification accuracy and rarely investigates the perception, reliance and trust of users~\cite{devillers2021human, picard200355}. One major challenge in improving these qualities is due to the lack of transparency to end-users, which is increasingly becoming more important in the context of practical affective computing research and application~\cite{tian2022aac, barrett2019emotional}. A lack of transparency with regards to how an integrated FER system uses the data input to predict the output and how it uses the predicted output in making a decision could affect the user's perception and reliance towards the system~\cite{linardatos2020explainable}. A system that is not transparent could be perceived as ambiguous, not accountable or even having mismatched goals with the user~\cite{abdul2018trends}.

\subsection{eXplainable Artificial Intelligence (XAI)}

In the field of artificial intelligence (AI), explainability (also sometimes equated with interpretability)~\cite{doshi2017towards, graziani2023global} aims to create a shared understanding~\cite{miller2019explanation} between end-users and the AI system that is being used. The main goal of these explainable AI systems or XAI is to make the inner workings of AI systems more interpretable and comprehensible to human users~\cite{lipton2018mythos, vilone2020explainable, vered2020demand}, provide greater transparency~\cite{miller2022we, gohel2021explainable}, help users gain greater confidence in model predictions~\cite{van2020interpretable} and engender appropriate trust in these systems~\cite{weld2019challenge}.

When investigating prevalent explanation methods, there are two main approaches: \textit{backpropagation-based} and \textit{perturbation-based} techniques~\cite{das2020opportunities}. In \textit{backpropagation-based} techniques such as Saliency Maps~\cite{simonyan2013deep}, the algorithm performs one or more forward passes through a network and generates attributions during the backpropagation stage. SHAP~\cite{lundberg2017unified} and LIME~\cite{ribeiro2016should} on the other hand, use \textit{perturbation-based} techniques in which the input instances are perturbed by occlusions, substitutions, masking, etc to generate an attribution representation which informs feature importance~\cite{das2020opportunities}. We choose these different explanation techniques as they are representative of the explanation methods extensively used currently in the field of XAI. While multiple comparison studies~\cite{aldughayfiq2023explainable, panati2022feature} have been conducted within other areas, these methods have yet to be compared within the context of FER.

\subsection{FER XAI}\label{subsection_fer_xai}
As discussed earlier, multiple XAI methods exist currently (such as Grad-CAM~\cite{selvaraju2017grad}, LIME~\cite{ribeiro2016should}, SHAP~\cite{lundberg2017unified}) which attempt to provide post-hoc explanations to the inner workings of different classifiers. However, despite these being powerful methods of explaining generic models in AI, they have been developed to work with \textit{any} classifiers and not 
specifically with the purpose of interpreting FER models. As is usually the case, generality can cause a loss of information which can affect performance. Prior comparison work mostly utilized these types of method categories to qualitatively assess the suitability of explanations~\cite{rathod2022kids, del2022understanding} which is subjective in nature. In other studies~\cite{ellingsen2022patient, lundberg2017unified}, authors use XAI techniques to determine which features the model places more importance on, so that re-training with critical data points can take place for better performance. Crucially, these studies do not empirically evaluate the perceptions and understanding of potential non-expert end-users who are important stakeholders in the use of this technology. Empirical experiments which conduct comparisons of end-user understanding and perception when provided with these different explanations have not been extensively researched in the context of FER to the best of our knowledge.  

We compare the performance of these widely used, \textit{general}, XAI methods to our novel explanation system (\textit{DEFAULTS}), tailored specifically for the task of FER. To develop our system, we leverage work from the Facial Action Coding System (FACS)~\cite{ekman1978facial}. The FACS encodes human facial expressions in terms of muscle activations on a person's face. Each muscle activation, or Facial Action Unit (FAU) represents a visible facial muscle movement that can give indications towards a person's emotional state. Examples of these FAUs include \textit{Eyebrow Raiser}, \textit{Nose Wrinkler}, \textit{Lips Part}, etc. \textit{DEFAULTS} is therefore grounded in formal theory with regards to human physiology when expressing facial emotions. 

In order to use FAUs as part of our explanations we first need to be able to detect them. To guarantee a fair comparison of model accuracy  we need to ensure that the comparable, \textit{general}, XAI methods are  able to generate explanations from the common reference FER model. To do this we leveraged prior work which predicted FAUs using various machine learning models, such as Support Vector Machines (SVMs)~\cite{baltrusaitis2018openface} or Deep Learning approaches~\cite{kim2019deep}. \citet{kim2019deep} developed a Convolutional Neural Network (CNN) which accepts input images and outputs a prediction of 8 emotion classes. Using the prediction vector and final convolutional layer, the authors fed that into a Deep Neural Network (DNN) which outputted a prediction vector on 15 classes of FAUs. This method allows for effectively generating FAU predictions using the DNN model. In our work we utilised the CNN model, which was used as the basis prior to the DNN predictor, as the common reference model over which we generated explanations using the other XAI methods described.

\section{Method}

\begin{figure}[]
    \centering
    \includegraphics[width=\linewidth]{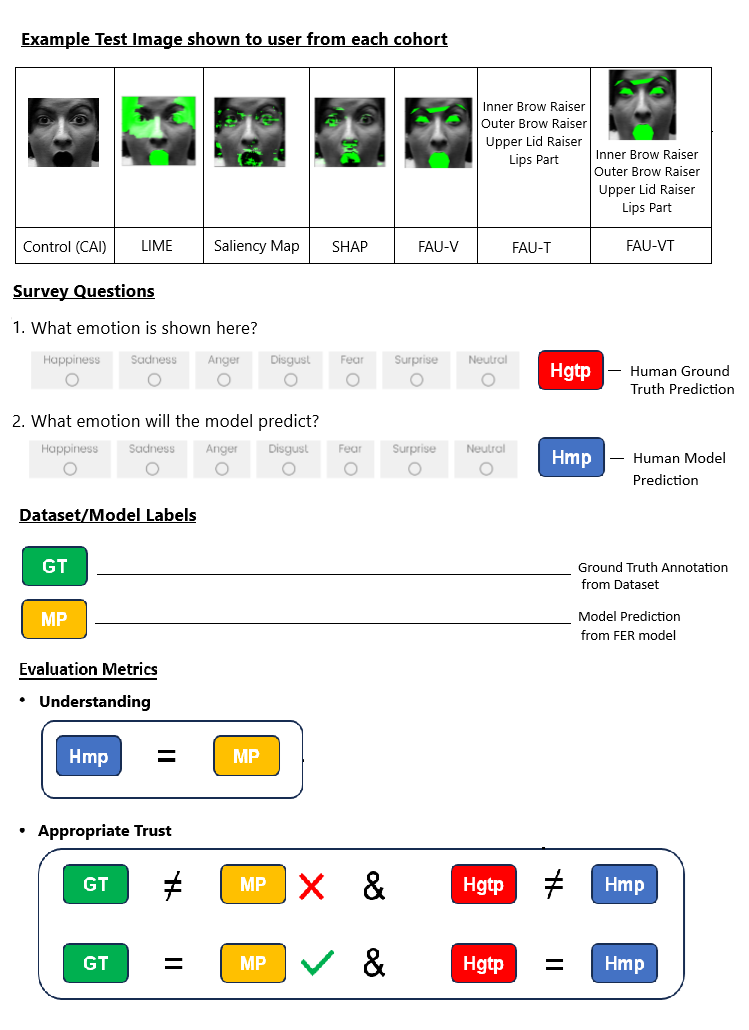}
    \caption{\label{overview_fig}An Overview Figure. (From top) Visualisation of example test image shown to each cohort. Survey Questions asked to users once they view the image and subsequent explanation image. Dataset model labels consisting of `GT' which is the annotated label in the dataset for this image and `MP' which is the FER model's actual prediction. Evaluation Metrics visualisation, as detailed in Section~\ref{evaluation_metrics_study2} (bottom).}
\end{figure}

An overview diagram of our study can be seen in \autoref{overview_fig}. The top section shows an example of images shown to participants of a person displaying an emotion. The first image shows how images were presented to the control group, with no additional explanations added to the image. The subsequent images show examples of explanations by each XAI method being compared.  
Below that we display the survey questions presented to participants after each test image. The questions in the test phase were developed to gauge participant understanding as well as level of trust towards the model's predictive capabilities. The answer to the first question establishes the \textit{human ground truth prediction} (HGTP), while the answer to the second question established the \textit{human model prediction} (HMP). 
We also include Evaluation Metrics (displayed at the bottom) to guide how we assess user understanding and trust. This evaluation is based on participant responses, dataset annotations, and the model's actual predictions. In the following subsections, we will provide more details on the FER model used as a baseline, the XAI methods chosen for comparison as well as our novel explanation method, the study design used, experimental procedure and finally, the participant selection and sample size determination.

\subsection{FER Model \& Explanation Methods}

We replicated the FER model developed by~\citet{kim2019deep}, as discussed in \autoref{subsection_fer_xai}, because it could be directly applied by both general XAI methods and FAU predictions. The CNN component of the model was trained on the CK+ dataset~\cite{lucey2010extended}, achieving a validation accuracy of 91.13\%. This CNN model took images as input and produced outputs in the form of a 1x8 vector, predicting the eight emotion classes. We used this CNN model to generate explanations with various state-of-the-art XAI methods. The methods chosen for comparison include LIME~\cite{ribeiro2016should} and SHAP~\cite{lundberg2017unified} which are representatives of state-of-the-art pertubation-based methods, as well as Saliency Maps~\cite{simonyan2013deep} which is a representative of state-of-the-art gradient-based method. By using the same CNN model to generate all explanations, we ensured a fair comparison as the accuracy of the model is kept constant. 

The DNN model was then trained as a separate head to the CNN model in order to predict FAUs. It used a concatenation of the final convolution layer of the CNN model (a 1x4032 Fully Connected node) and the 1x8 emotion prediction vector from the CNN to predict the FAUs present in an image. The DNN model outputs a 1x15 vector of boolean active/inactive predictions for FAUs. When cross-validated using the CK+ dataset, this model achieved a minimum accuracy of 96.33\%, similar to the CNN model. With this in place, we now have the ability to generate \textbf{FAU-based} explanations from the previously trained CNN model.

\begin{figure}[H]
    \centering
    \includegraphics[width=\linewidth]{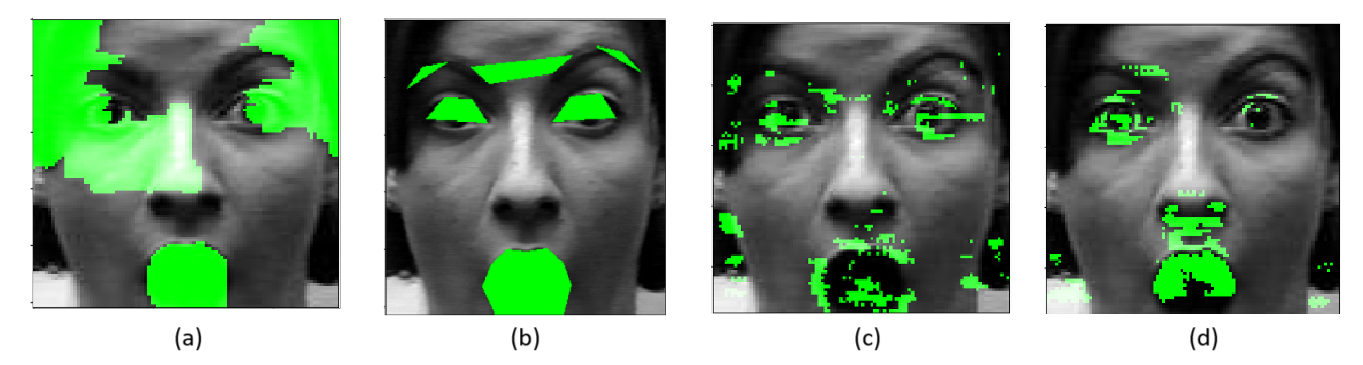}
    \caption{\label{xai_standard_label}Standardized masked image explanations generated by (a) LIME, (b) FAU-based, (c) Saliency Map and (d) SHAP}
\end{figure}

The output of the DNN model, a 1x15 vector of boolean active/inactive FAUs, could now be converted directly into textual explanations of activated FAUs (such as \textit{Eyebrow Raiser}, \textit{Nose Wrinkler}, \textit{Lips Part}, etc). An example of this can be seen in the Example Test Image section in \autoref{overview_fig} under the \textbf{FAU-T} method. We then sought to develop a method to generate visual representations of FAUs to complement these textual explanations. Utilizing the work from~\cite{perveen2020configural}, we identified the facial landmarks associated with each FAU activated. We then drew contours between the points to represent our finalized visual FAU explanations. An example of this can be seen in the Example Test Image section in \autoref{overview_fig} under the \textbf{FAU-V} method. Both the FAU-T and FAU-V methods were combined to generate the final \textbf{FAU-VT} method as seen in \autoref{overview_fig} which represents explanations generated by \textit{DEFAULTS}. The standardized masked representations presented to participants are shown in \autoref{xai_standard_label}. The decision to use standardized masked representations ensured that each representation shown did not bias the user. For each masked image, the original image was also displayed to users for easy comparison of the regions highlighted by the mask.

\subsection{Conditions}

The experiment was conducted as a between-subject study with seven conditions. The list below outlines the seven cohort groups, which were based on the explanation methods described in the previous subsection. Example images generated for each cohort can be seen in the Example Test Images section in \autoref{overview_fig}.

\begin{enumerate}
    \item \textbf{CONTROL (CAI)} Participants were not provided with any explanations. 
    \item \textbf{LIME} Participants were provided with explanations using the LIME method.
    \item \textbf{SALMAP} Participants were provided with explanations using the Saliency Map method.
    \item \textbf{SHAP} Participants were provided with explanations using the SHAP method.
    \item \textbf{FAU-T} Participants were provided with explanations using only the textual modality of \textit{DEFAULTS}. Example: \textit{Inner Brow Raised, Lips Parted, Nose Wrinkled, etc}.
    \item \textbf{FAU-V} Participants were provided with explanations using only the visual modality of \textit{DEFAULTS}.
    \item \textbf{FAU-VT} Participants were provided with explanations using both the textual and visual modality of \textit{DEFAULTS}.
\end{enumerate}

\subsection{Research Questions \& Hypotheses}

The goal of this work is to identify which types and modalities of explanations enhance users' understanding of FER models and foster appropriate trust. We hypothesized that users who receive both textual and visual FAU-based explanations would better understand the system, demonstrated by their ability to more accurately predict the model's outputs compared to users who receive no explanations at all. We further hypothesized that this increased understanding will also engender more appropriate trust in the model. Based on this, we explored the following research questions and associated hypotheses.

Research Questions:
\begin{enumerate}
    \item RQ1: \textit{Which explanation method most effectively helps users understand how the model works in practice?}
        \subitem H1.1: Users provided with visual explanations based on FAUs will show a stronger alignment with the model's predictions compared to users provided with other types of visual explanations.
        \subitem H1.2: Users provided with explanations in both visual and textual modalities will show a stronger alignment with the model's predictions compared to users provided with only one modality of explanation.
    \item RQ2: \textit{Which explanation method helps users develop the most appropriate level of trust in the model?}
        \subitem H2.1: Explanations based on visual FAUs will foster  more appropriate perceived and demonstrated trust in the model, compared to other types of visual explanations.
        \subitem H2.2: Explanations provided as both visual and textual modalities will lead to more appropriate perceived and demonstrated trust in the model, compared to individual modalities.
\end{enumerate}

\begin{table}[h!]
\centering
\begin{tabular}{|l|p{2.8cm}|p{3.3cm}|}
\hline
\textbf{Part \#} & \textbf{Analyses} & \textbf{Cohort Groups} \\ \hline
1 & Explanation Types (visual only) & Control (CAI), LIME, SALMAP, SHAP, FAU-V \\ \hline
2 & Explanation Modality & Control (CAI), FAU-T, FAU-V, FAU-VT \\ \hline
\end{tabular}
\vspace{10pt}
\caption{Analyses}
\label{analysis_parts_table}
\end{table}

In order to answer Research Question 1 and 2 the following analyses were created, as shown in \autoref{analysis_parts_table}. Analysis Part \#1 (Explanation Types (visual only)) answers H1.1 and H2.1 whereas Analysis Part \#2 (Explanation Modality) answers H1.2 and H2.2.
The design of both experiment analyses was captured in a pre-registration on Open Science Framework\footnote{Links to pre-registration: \url{osf.io/2ez64, osf.io/35upf}} which was made public before the experiment commenced. The study procedure was reviewed and approved by the Monash University Human Research Ethics Committee under Project ID 37086.

\subsection{Experiment Procedure}

\begin{figure}[h!]
    \centering
    \includegraphics[width=\linewidth]{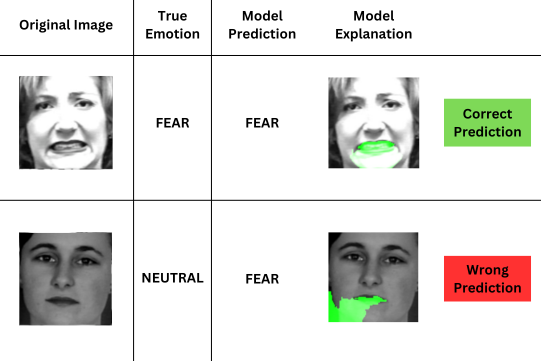}
    \caption{~\label{lime_train}Example images shown for the LIME cohort during the training phase.}
\end{figure}

All cohorts gave their consent at the start of the session and completed a brief pre-questionnaire on demographics. Participants then entered a training phase specific to their cohort. The images selected for the training phase were randomly sampled from the CK+ dataset. During training participants in the control cohort viewed images with facial expressions and were informed of both the ground truth annotated emotion and the model's prediction. These participants did not receive any explanations. Participants in the other six explanation cohorts were shown the original images, ground truth annotation and model prediction as well as an explanation of the model according to their cohort group. For each of the 7 Ekman emotion categories~\cite{ekman1978facial} (neutral, anger, sadness, happiness, fear, surprise and disgust) participants were shown one example of an accurate model prediction and one example of an inaccurate model prediction (an example for LIME is shown in \autoref{lime_train}).  Participants were explicitly informed that the examples of accurate and inaccurate predictions were not reflective of the model's overall accuracy.

Following the training phase, participants moved on to the testing phase, which included a total of 28 images (4 images for each of the 7 emotion classes). The images were randomly sampled from the Aff-Wild2 dataset~\cite{kollias2018aff} which features non-posed expression images under various lighting conditions, age groups, and genders. A balanced distribution of 50\% correct and 50\% incorrect prediction images was chosen to reflect state-of-the-art FER accuracy of 52\% on this dataset~\cite{ryumina2022search}. Therefore, for each emotion class 2 correct and 2 incorrectly predicted images were selected to be tested. By using state-of-the-art FER benchmark with in-the-wild data to determine the accuracy of the model to be presented to users, our approach does not rely on the underlying model's accuracy offering insights into the application of XAI in realistic FER contexts.

Similar to the training phase, for each image in the testing phase, participants were shown the original test image along with  explanations of the model based on their specific cohort (either LIME, SHAP, SALMAP, FAU-T, FAU-V or FAU-VT) (\autoref{overview_fig}, top). Participants in the CONTROL (CAI) cohort were shown only the original test image without any explanations. Model predictions were not displayed to participants during the test phase. The order in which the images were presented was randomized to control for any potential ordering effects.

\begin{figure}[h!]
    \centering
    \includegraphics[width=\linewidth]{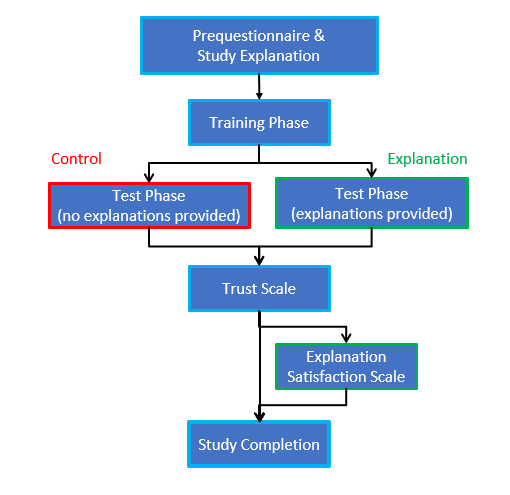}
    \caption{~\label{processflow}Process Flow of Experiment Procedure.}
\end{figure}

Participants were then asked two questions (as shown in the \textit{Survey Questions} section of \autoref{overview_fig}). For both questions, participants were required to select one option from the seven emotion categories. The seven emotion categories were used instead of the eight categories the model was trained on because the test dataset (Aff-wild2) does not include the ``contempt'' emotion. After completing the 28 images, participants answered the Trust Scale and Explanation Satisfaction scale~\cite{hoffman2018metrics}. The survey items were designed to assess the participant's perceived trust in the FER model after being exposed to explanations, as well as to measure the helpfulness of the explanations in the context of FER. Participants in the CONTROL cohort were not asked to complete  the Explanation Satisfaction Scale, as no explanations were provided to them. A full process flow of the experiment procedure can be seen in \autoref{processflow}.

\subsection{Evaluation Metrics}\label{evaluation_metrics_study2}

To evaluate and address RQ1 and RQ2, the following metric items were used:

\begin{enumerate}
        \item \textbf{Human Model Prediction (Hmp) Accuracy}: The number of times that participants answered Question \#2 from the Survey Questions section in \autoref{overview_fig} correctly out of the total number of questions. This measured a participant's ability to guess the model's prediction, which contributes towards answering RQ1 H2.1 and H2.2.
        \item \textbf{Appropriate Trust}: This contributes towards answering RQ2 H3.1 and H3.2. In \autoref{overview_fig}, ``GT'' represents the ground truth annotation label from the Aff-Wild2 dataset and ``MP'' represents the model prediction of our FER model. Trust is considered appropriate when either:
        \begin{enumerate}
            \item For the correct test images (where GT is the same as MP): Participant answers for Question \#1 (Hgtp) and Question \#2 (Hmp) from the Survey Questions section in \autoref{overview_fig} are the same. This means that participants trusted that the model will predict the same emotion as what they answered for Question \#1. OR;
            \item For the incorrect test images (where GT is not the same as MP): Participant answers for Question \#1 (Hgtp) and Question \#2 (Hmp) from the Survey Questions section in \autoref{overview_fig} were different. This means that participants did not trust that the model will predict the same emotion as what they answered for Question \#1.
        \end{enumerate} 
\end{enumerate}

Other metrics captured and analysed which do not directly relate to the research questions are shown below. These metrics help to inform the study and add more context to the discussion.
\begin{enumerate}
    \item \textbf{HP Accuracy}: The number of times that the participant answered Question \#1 from the Survey Questions section in \autoref{overview_fig} correctly (i.e., Hgtp = GT) out of the total number of questions. This measured a participant's ability to recognise facial expressions.
    \item \textbf{Trust Scale Overall Score}: This was calculated as the sum of the points for each question in the scale after deducting the score from item \#6 (\textit{``I am wary of the emotion recognition system.''}) which was negatively weighted. 
    \item \textbf{Explanation Satisfaction Scale Overall score}: calculated as the sum of the points for each question in the scale.
\end{enumerate}

We conducted a one-way ANOVA test for each of the metric items shown above to compare the means of multiple groups within each analysis part. If the ANOVA test revealed any significance, we conduct a subsequent post-hoc Tukey HSD test to determine which groups were significantly different from one another.

\subsection{Participants}

A G*Power analysis was conducted to determine the number of samples required for each participant group. The goal was to obtain 0.8 power at the standard .05 alpha error probability. Using a one-way ANOVA test as our planned statistical test and an estimated medium effect size of 0.25, we found the total sample size recommended to be 200 participants for Analysis \#1 which has 5 groups. For Analysis \#2, the total sample size recommended was 160 participants for a medium effect size of 0.265 and 4 groups. Due to the overlap of the Control and the FAU-V cohorts for both analysis parts, the number of participants recruited was 280 in total.

Participants were recruited from the Prolific platform which has been shown to produce high quality responses~\cite{douglas2023data, eyal2021data}. Participants were selected from English speaking countries (Australia, UK, US, Canada) between the ages of 18 to 100 years old with the male:female gender ratio balanced at 50\%. Participants were paid at a rate of £6 per hour and the study's estimated time was 30 minutes based on a pilot study conducted with 21 participants. 

To ensure the quality of responses, two attention checks were put in place within the survey which tested participants based on their ability to recognize generic objects. The questions were not relevant to the test questions presented in the main body of the survey. Participants who completed the survey in less than half of the median time (approximately 30 minutes) and failed both attention checks were removed from the study results to ensure only quality responses were captured.

\section{Results}\label{study_results}

We now describe the results of our study. We use boxplots to present our results, in which the top line of the box represents the 3rd quartile and the bottom line represents the 1st quartile of the data points. The middle ``bolded'' line in the box represents the 2nd quartile or median of the data. Dots in the boxplots represent outliers in the data. A one-way ANOVA was performed for both analyses (referred to in \autoref{analysis_parts_table}) for each of the metric items discussed in \autoref{evaluation_metrics_study2}. The ANOVA test helps to determine how the means of the groups differ with regards to the evaluation metrics through different explanation types. If significant effect is found, this means that there is a significant difference in the means of at least two groups. However, the ANOVA test alone cannot determine which two groups differ. Therefore, for results with significance, a Tukey HSD test was then carried out to determine which groups were significantly different as this test compares all pairwise differences for significance.

\subsection{FER model understanding}
The one-way ANOVA test revealed no significance between cohorts in the Explanation Type part ($F(4, 195) = 0.8, p = 0.53$) as well as the Explanation Modality part ($F(3, 156) = 1.02, p = 0.39$) for the Human Prediction (HP) Accuracy. The HP Accuracy was determined by calculating the number of times the participant got Question \#1 correct (Hgtp) in the Survey Questions shown in \autoref{overview_fig}. This shows that participants in this study were similar in their ability to recognize emotions and indicates that no specific group had better emotion recognition abilities which could be seen as an advantage over other groups.

\begin{figure}[h!]
    \centering
    \includegraphics[width=\linewidth]{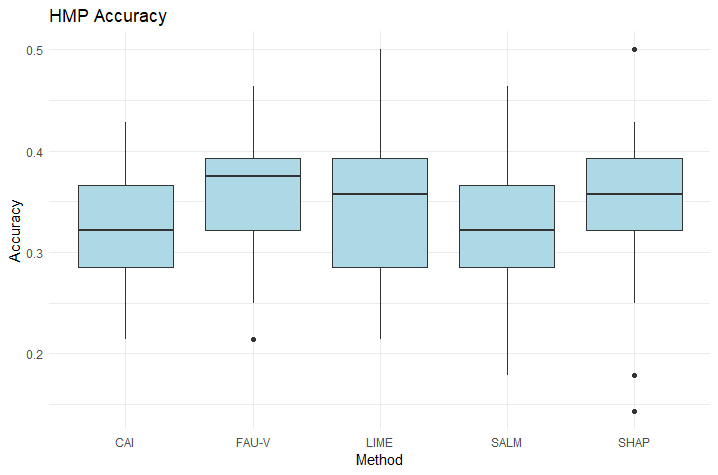}
    \caption{\label{hmp_accuracy_method_img}Hmp Accuracy between groups with different XAI methods. No significance was found between the groups.}
\end{figure}

\begin{figure}[h!]
    \centering
    \includegraphics[width=\linewidth]{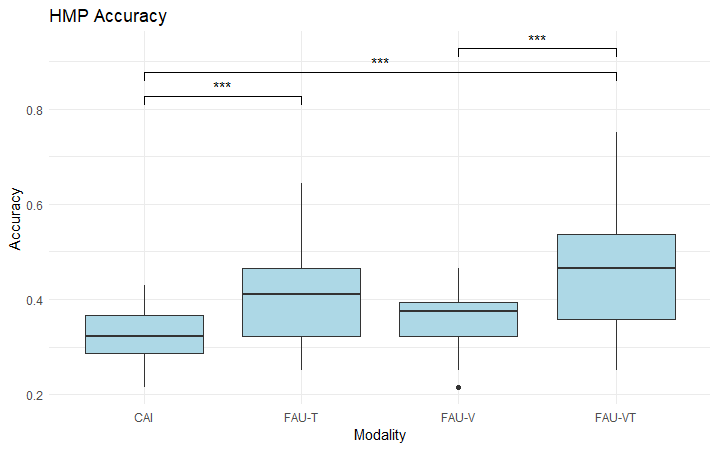}
    \caption{\label{hmp_accuracy_modality_img}Hmp Accuracy between groups with different XAI modality. The FAU-VT method has higher Hmp accuracy than FAU-V and Control. FAU-T has higher Hmp accuracy than Control.}
\end{figure}

In order to answer \textbf{H1.1} and \textbf{H1.2} for Research Question 1 (RQ1), we analysed the Human Model Prediction (Hmp) Accuracy. The Hmp Accuracy was determined by calculating the number of times the participant got Question \#2 correct (Hmp) in the Survey Questions shown in \autoref{overview_fig}. The one-way ANOVA test revealed no significance between cohorts in the Explanation Type part ($F(4, 195) = 2.21, p = 0.07$) as can be seen in \autoref{hmp_accuracy_method_img}. However, the test did show a large significance for the Explanation Modality part ($F(3, 156) = 15.41, \textbf{p < 0.001}$) as shown in \autoref{hmp_accuracy_modality_img}. A post-hoc Tukey HSD test indicates that the means for the FAU-T and FAU-VT groups was significantly higher ($\textbf{p = 0.0002 \& p < 0.0001}$) than the CONTROL group. The test also showed that the mean for the FAU-VT group was significantly higher ($\textbf{p = 0.0001}$) than the FAU-V group. 

\subsection{Appropriate trust of the FER model}
In order to answer \textbf{H2.1} and \textbf{H2.2} for Research Question 2 (RQ2), we then analyse the Appropriate Trust Score for each analysis. As mentioned earlier in \autoref{evaluation_metrics_study2}, trust is considered appropriate when the participant answers that their guess of the emotion shown on the image and the emotion the model will predict is the same for correct images (where the model prediction is the same as the ground truth). Alternatively, trust is also considered appropriate when the participant answers that their guess of the emotion shown on the image and the emotion the model will predict is different for incorrect images (where the model prediction is different than the ground truth).

\begin{figure}[h!]
    \centering
    \includegraphics[width=\linewidth]{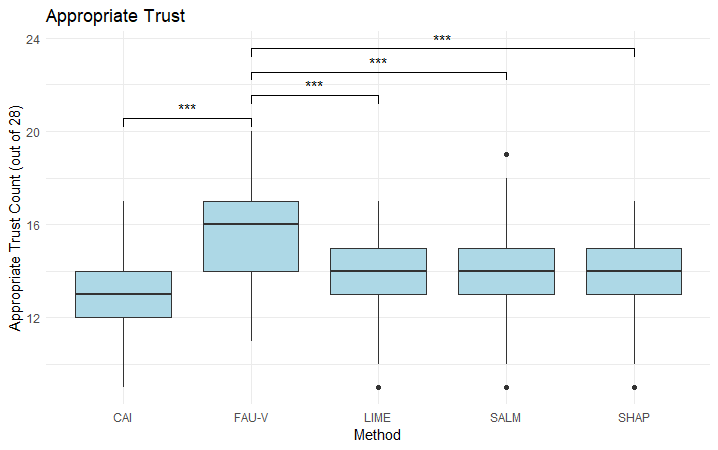}
    \caption{\label{appr_trust_method_img}Appropriate Trust between groups for Method Type (visual-only) Part. FAU-V type has higher appropriate trust as compared to all other visual-only types and the Control cohort.}
\end{figure}

\begin{figure}[h!]
    \centering
    \includegraphics[width=\linewidth]{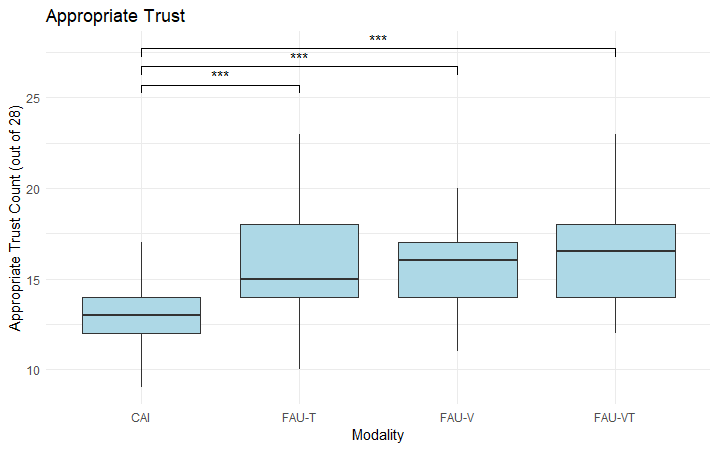}
    \caption{\label{appr_trust_modality_img}Appropriate Trust between groups for Modality Part. FAU-based methods have higher appropriate trust as compared to the control cohort.}
\end{figure}

The one-way ANOVA test revealed significance between cohorts in the Explanation Type part ($F(4, 195) = 11.65, \textbf{p < 0.0001}$) (seen in \autoref{appr_trust_method_img}) as well as the Explanation Modality part ($F(3, 156) = 17.26, \textbf{p < 0.0001}$) (seen in \autoref{appr_trust_modality_img}). For analysis part 1 (Explanation Type), the post-hoc Tukey HSD test showed that the FAU-V method engendered significantly higher ($\textbf{p-value < 0.001}$) appropriate trust when compared to the other visual-only explanation methods (LIME, Saliency Map, SHAP) and the Control cohort. For analysis part 2 (Explanation Modality), the test showed that all the FAU modalities engendered significantly higher ($\textbf{p-value < 0.0001}$) appropriate trust as compared to the control cohort.  

We excluded analytic graphs of the Trust Scale and Explanation Satisfaction Scale outcomes as no significance was found with this self-reported questionnaire method.

\section{Discussion}
In this study we sought to understand if XAI methods can help end-users understand \textit{what} a FER model will predict and \textit{when} to trust an FER model's prediction. We proposed a novel FAU-based XAI method (\textit{DEFAULTS}) and compared it with current XAI methods. We first compared different XAI methods using visual-only explanations.  Second, we compared different modalities, namely visual and textual, within the proposed method itself.

\subsection{Improving user understanding of an FER model with visual and textual explanations}
When considering Research Question \#1, which investigated the effect of explanations in terms of model understanding, the Hmp Accuracy results did not support \textbf{H1.1}, which hypothesised that participants shown the FAU-based visual explanations would have higher alignment with the model's prediction as compared to other forms of visual explanations. No significant difference was found using the one-way ANOVA test and thus this hypothesis was rejected. One possible reason for this could be that the information relayed by the FAU-V method was purely provided through visual aid, similar to the other XAI methods, and that purely visual-only techniques are not sufficient to educate end-users on the knowledge required to determine what the model could predict. This is a critical finding as most explanations generated on adapted XAI techniques on FER systems centers around providing visual-only explanations which are clearly not significant enough to improve user understanding of a FER model.

However, for the Hmp results in analysis \#2 which investigates \textbf{H1.2}, we see that the Hmp Accuracy was significantly higher with the FAU visual+textual method (FAU-VT) compared to the FAU visual-only method (FAU-V) and the control cohort. Interestingly, the FAU textual-only method (FAU-T) also had a significantly higher Hmp Accuracy compared to the control cohort as well. This results partially support \textbf{H2.2}: FAU methods using both visual+textual and the textual-only modalities are better than no explanations in helping the user understand the model better. The FAU-VT method shown as having a higher score than the FAU-V method also indicates that combining the textual modality with visual enhances its performance and increases user understanding of the model and subsequent prediction. Therefore, since FAU-based methods seem to lead users to understand the model better, we find that the textual component of explanations are extremely important but the combination of both textual+visual would be the best in terms of improving user understanding. This is an important finding as, to the best of our knowledge, there are no other existing FER explanation models that adopt this approach. 

\balance

\subsection{Improving appropriate user trust in FER model with FAU-based explanation}
Next we consider Research Question \#2 which investigates which explanation method and modality helps to engender appropriate user trust in the model. The results, as shown in \autoref{appr_trust_method_img}, indicate that FAU-V engenders significantly higher appropriate trust compared to the control cohort and other visual-based explanation cohorts. This confirms \textbf{H2.1}: visual based FAU explanations can engender higher appropriate demonstrated trust compared to other forms of visual explanations. This significance was not found in the investigation into user understanding, indicating that FAU visual methods may be helpful in providing a certain ``intuition'' to users on whether or not a model is behaving well or misbehaving. However, the visual-only FAU method is not enough to really improve user understanding towards the inner working of the model. 

In terms of the different modalities and their effect on appropriate trust levels, we refer to the results shown in \autoref{appr_trust_modality_img}. The results indicate that although each different modality within the FAU-based method had a significantly higher appropriate trust as compared to the control cohort, there was no significance between the different modalities itself. This suggests that, as a whole, the FAU-based methods are more successful at engendering appropriate trust as compared to not having explanations at all. Users who were provided with this explanation type had a better understanding of when the model will succeed or fail.

In analysing these results we can see that, in general, variants of the FAU-based method are better at improving user understanding or at least intuition when a model is performing well and when it is not, with the textual component of FAU-based explanations indicating a better improvement of user understanding. Furthermore, the combination of textual and visual explanation modalities significantly improves user knowledge of the model compared to each modality individually. This aligns directly with our hypothesis and initial thoughts where explanations on FER using FAUs are essentially grounded in human physiology while also being formal methods of explaining. While formal explanations guarantees soundness of explanations, it also reduces the amount of redundant information which could confuse users that interact with these types of systems~\cite{marques2022delivering}. 

By focusing on explainable techniques that offer a practical and naturalistic account of understanding towards an AI model~\cite{paez2019pragmatic}, there is definitely room for improvement in improving transparency and understanding~\cite{tian2022aac} towards these AI models for end-users. However, the lack in significant difference between each cohort with regards to the participant's self-reported outcomes of the Trust Scale and Explanation Satisfaction Scale shows that much work is still needed to measure and improve user's trust and satisfaction of explanations in general. Ultimately, these insights and improvements to explanations should culminate in improved perception, reliance and trust of users~\cite{devillers2021human, picard200355, linardatos2020explainable} which enable fair use of FER technology through raising awareness of the potential biases and limitations of current FER systems~\cite{devillers2023ethical, devillers2021human}.

\subsection{Limitations}
Our work should be viewed in light of the following limitations. One limitation of our work is that the experiments were conducted with a single type of stakeholder (laypeople) and within demographics which speak English as a primary language. This limits our scope as we would not know how this explanation types affect user understanding and appropriate trust when tested on a more multicultural and multilingual group with diverse professional backgrounds. We also plan to extend our work to a more interactive setting with some element of risk involved as this would be a more practical way of measuring trust towards a system. Furthermore, the textual component of FAUs denote some sort of ``action'' whereas the visual component only relay static information. This could be perceived as a limitation of comparing the visual and textual representations as it may not be directly comparable. However, this presents a clear advantage of using textual FAU explanations as they provide more contextual information compared to static images. In future comparisons, we plan to expand the visual representation with animation to add an ``action'' to visual explanation similar to the textual component. Last but not least, although we aimed for a representative of each XAI types, we could not test all state-of-the-art XAI methods which are currently being developed. In the future, we aim to provide more literature on XAI methods to see their relevance to explaining FER models and develop benchmarks for evaluating XAI for FER.

\section{Conclusion}

We proposed \textit{DEFAULTS}, a novel explanation method specifically designed for facial expression recognition (FER), and evaluated its effectiveness in comparison with state-of-the-art XAI methods.  We conducted an online crowd-sourcing experiment to investigate how different explanation methods influence  user understanding and appropriate trust in the model. Our findings revealed that, compared to participants without explanations, those provided with the \textit{DEFAULTS} method using both visual and textual Facial Action Unit (FAU)-based explanations, had a better understanding of the model. This was reflected in their higher accuracy in predicting the model's outputs, as well as their more appropriate trust in the model's accuracy - trusting the FER model when it was accurate and withholding trust when it was inaccurate. In contrast, existing XAI methods (LIME, SHAP, Saliency Map) showed no significant improvement in user understanding or appropriate trust. We also demonstrated that purely visual explanations for FER models are insufficient to enhance user understanding of how the model operates. Our findings suggest that FAU-based methods offer greater intuition for end-users to assess whether the model is performing well or misbehaving. This work is the first to explore effective XAI methods and explanation modalities for FER models, highlighting the benefits of generating multimodal explanations grounded in emotion theories to improve user understanding and appropriate trust in FER systems. Our research thus lays the foundation for the future development of more explainable emotion-aware systems.






\bibliographystyle{ACM-Reference-Format} 
\bibliography{bibliography}


\end{document}